\documentclass[twocolumn,pre,nofootinbib,floatfix,superscriptaddress,showkeys]{revtex4-1}

\usepackage{hyperref}
\usepackage{dcolumn}
\usepackage{graphicx}
\usepackage{rotating}
\usepackage{amsmath,amsfonts,amssymb}
\usepackage{multirow}
\usepackage{color}
\usepackage{filecontents}

\begin{document}
\title{Predicting the outcomes of policy diffusion from U.S. states to federal law}

\author{Nora Connor}
\email{nora.connor@colorado.edu}
\affiliation{Department of Computer Science, University of Colorado, Boulder CO, 80309 USA}

\author{Aaron Clauset}
\email{aaron.clauset@colorado.edu}
\affiliation{Department of Computer Science, University of Colorado, Boulder CO, 80309 USA}
\affiliation{BioFrontiers Institute, University of Colorado, Boulder CO, 80303 USA}
\affiliation{Santa Fe Institute, Santa Fe NM, 87501 USA}

\begin{abstract}
In the United States, national policies often begin as state laws, which then spread from state to state until they gain momentum to become enacted as a national policy. However, not every state policy reaches the national level. Previous work has suggested that state-level policies are more likely to become national policies depending on their geographic origin, their category of legislation, or some characteristic of their initiating states, such as wealth, urbanicity, or ideological liberalism. Here, we tested these hypotheses by divorcing the set of traits from the states' identities and building predictive forecasting models of state policies becoming national policies. Using a large, longitudinal data set of state level policies and their traits, we train models to predict (i) whether policies become national policy, and (ii) how many states must pass a given policy before it becomes national. Using these models as components, we then develop a logistic growth model to forecast when a currently spreading state-level policy is likely to pass at the national level. Our results indicate that traits of initiating states are not systematically correlated with becoming national policy and they predict neither how many states must enact a policy before it becomes national nor whether it ultimately becomes a national law. In contrast, the cumulative number of state-level adoptions of a policy is reasonably predictive of when a policy becomes national. For the policies of same sex marriage and methamphetamine precursor laws, we investigate how well the logistic growth model could forecast the probable time horizon for true national action. We close with a data-driven forecast of when marijuana legalization and  ``stand your ground'' laws will become national policy.\end{abstract}


\maketitle

\section{Introduction}
In the United States, many state laws are not the result of independent innovations. Instead, they arise from the diffusion of ideas and policies from one state to another \cite{karch2007emerging, gray1973innovation}. Because individual states have a degree of autonomy, they can both independently implement and evaluate policies as well as copy policies enacted by another state. If the policy works well, it may spread to other states or even be enacted at the national level as a federal policy. In this way, states act as independent ``laboratories of democracy'' \cite{new_state_ice_co_v_liebmann}, allowing for experimentation while also facilitating the spread of policies that work well. The process by which policies spread across the states is sometimes called ``policy diffusion.''

Policy diffusion is generally defined as any process in which a government body makes policy decisions predicated on the previous policies implemented in other jurisdictions \cite{dobbin2007global}. Much of the research on policy diffusion focuses on U.S. policies, because U.S. states are semi-autonomous and the state and federal governments have a relationship of parity (i.e., the federal government and the state governments have equal status under the law and neither is subordinate to the other) \cite{walker1969diffusion, boushey2010policy, berry1990state}. In addition, there is a long historical record of policies and their spread across states and between these levels of government from which to draw insights.

Many studies of policy diffusion have focused on disentangling different social mechanisms that can drive or limit diffusion, such as imitation, coercion, or socialization. A weakness of these studies, however, is their focus on a single policy and its spread \cite{desmarais2015persistent, jansa2018copy}. Boushey \cite{boushey2010policy} argues that these single-policy studies arise due the prevalence in American politics of event history analyses, introduced by Berry \& Berry \cite{berry1990state}). These analyses model the spread of a single policy by the regression of a set of variables that can change across time for each jurisdiction, such as fiscal health, religious fundamentalism, and frequency of interstate communications \cite{boehmke2009approaches}. Graham \textit{et al.} \cite{graham2013diffusion} suggest that comparing multiple policies, across a range of categories, would allow us to discover some of the variables that impact diffusion more broadly (as in Ref.~\cite{walker1969diffusion}), though little work in this direction has been carried out. Event history modeling can be useful at the level of individual states to understand the contextual details of particular policies but ultimately says little about the underlying variables that drive diffusion more broadly.

Past studies also rarely consider data-driven forecasts of future diffusion of particular policies. Instead, research on policy diffusion has been largely retrospective and descriptive, aimed at understanding past diffusion and policy adoption \cite{boushey2010policy, berry1999innovation}. This work has produced several hypotheses about factors that shape the subsequent diffusion dynamics of policies. Specifically, past work suggests that states with more wealth \cite{walker1969diffusion, jansa2018copy}, higher urban density \cite{walker1969diffusion, shipan2012policy}, longer and more frequent legislative sessions \cite{shipan2006bottom}, certain geographic location \cite{mooney2001modeling}, and a more liberal ideology \cite{boushey2010policy} are predicted to be more innovative, meaning more likely to produce policies that spread. 

Here, we take a different approach, using predictive models that incorporate features of each policy to make predictions about how quickly and how far a policy will spread. This approach bears some similarity to recent efforts to predict the size or growth of information sharing ``cascades'' in social media, as in \cite{martin2016exploring} and \cite{cheng2014can}, and to characterize the spread of ideas in academia \cite{morgan2018prestige}. Here,  we combine several data sets of state-level policies, which we extended to include data on whether each policy ultimately passed at the national level and when. Building on the hypotheses from previous policy diffusion studies, we assign factors to each policy that represent its early adopters' wealth, urbanicity, innovativeness, legislative professionalism, and ideology. We then build a predictive model of whether a given state policy would become a national law given its associated factors. For policies that become national policies, we develop a second model for predicting the threshold number of states required to become national law as a function of its associated factors. Hence, the combination of these two models enables us to characterize the degree to which a national policy outcome is predictable from its associated factors. Finally, we develop a third model based on simple logistic growth that generates a statistical forecast of when a policy will become national. We find that hypothesis-based covariates predict national enactment only slightly better than chance, and that a fixed threshold of approximately 25 states outperforms more sophisticated models for a threshold of national action. Finally, in cases where there was federal action, our logistic growth model accurately forecasts the approximate year of national action with high accuracy across several well-known policies. We then use this model to  make specific forecasts for future federal actions for two policies that are currently spreading today.

\subsection*{Modeling policy diffusion}
By using a predictive modeling approach to learn which policy characteristics correlate with greater or faster diffusion, we can quantitatively assess existing hypotheses about policy diffusion \cite{boushey2010policy, berry1999innovation} and identify new patterns. For instance, we might expect that policies relating to child safety diffuse widely and quickly, or that the rate of diffusion has increased over time due to the ease of internet communication, or because of Political Action Committees (PACs) and non-profit organizations that promote partisan policies across state legislatures, e.g., organizations like the American Legislative Exchange Council and the State Innovation Exchange \cite{greenblattarticle, drutman2014evaluating}.

For policies that have not yet become law via a federal action, the cumulative number of states having adopted the policy often resembles a logistic growth function up to a maximum of 50 states. This S-shaped curve represents an archetypical pattern in which a policy is adopted slowly by a small number of states, then gains momentum as it spreads to many more states, and finally the spread tapers off as the remaining states adopt the policy and it approaches the 50-state limit \cite{boushey2010policy, Warnerarticle, gray1973innovation}. This S-shaped curve can be interrupted, however, if a national policy is implemented in the middle of this diffusion process. Effectively, such an event instantaneously moves the count to all 50 states. This disruption represents a change point, at which the policy spreading process abruptly changes from a bottom-up to a top-down mechanism. From a modeling perspective, the number of states that had adopted the policy just before the change point can be viewed as a kind of threshold.

We divide this overall diffusion process into three parts, which we model separately. First, will a given policy become national law? Second, if so, what is the threshold after which it becomes a national law? And finally, given a currently spreading policy, how far will it spread?

For the first question, we develop a model that makes a binary prediction about whether a state-level policy will become a federal policy by triggering a national action. To learn this model, we use a set of policy covariates that represent a broad range of policy characteristics, such as its broad policy category and the era in which it was first introduced, as well as characteristics of the policy's early adopters, such as state wealth, urbanicity, ideology, and legislative professionalism. Among the policies that did ultimately yield national laws, we focus on modeling a process by which policies undergo federal action and became federal law prior to diffusing nationally to all 50 states. For our models, all federal actions are considered to be equivalent, whether a policy is decided in a federal court case (e.g., Obergefell vs. Hodges, which legalized same sex marriage), passed via legislation in Congress (e.g., Katie's Law, which funds enhanced DNA collection for felony arrests), or enacted by an executive order from the President (e.g., Deferred Action for Childhood Arrivals or ``DACA'' policy, which allowed children of immigrants who were brought to the US to stay in the country legally for renewable two-year periods).

For the second question, we develop a model to predict the threshold number of states at which a policy will be nationally adopted. To learn this model, we use the same set of policy covariates from the first question. However, we limit the data to the subset of policies that triggered or resulted in a federal adoption, and use these positive examples of national action to predict their corresponding thresholds of the fraction of states required for that action.

Finally, for the third question, we test whether the two previous models, when combined with a parametric S-curve model for the temporal evolution of the number of states that have adopted the policy, is able to correctly predict the long-term spread and adoption of a policy. Specifically, we test the predictability of the long-term outcome of policy diffusion, as a function only of its early diffusion dynamics. To make this a realistic forecasting test, we restrict the training data to the policy's covariates and the adoption times of the first 5 or 10 states. To better model the underlying statistical uncertainty in these forecasts, we use a bootstrap approach to estimate an ensemble of models that collectively describe the underlying variance in the spreading process. The outcome of this approach is a distribution of years of when a particular policy would become national law. The broader this forecast distribution, the more uncertain the future trajectory of the diffusion process.

It is important to point out that none of our models incorporate the specific identity or geographic location of states in the data. As a result, we treat states as independent and identical entities, which simplifies the models and analyses, and improves the efficiency with which we can learn from small data sets. Accounting for factors associated with specific state identities, regional trends, state innovativeness scores, etc., which may require larger data sets, is left for future work. 

Understanding which policies spread well or not, and which policies tend to produce federal action, will shed new light on the dynamics of governmental policies and how ideas spread between state governments. Our results may also inform the work of so-called ``go-betweens,'' which are policy advocates or lobbyists who work across jurisdictional lines, such as think tanks and non-profits, by identifying the characteristics of policies that are more likely spread \cite{graham2013diffusion}. The degree to which policy spreading dynamics are predictable versus unpredictable, and the covariates that correlate with either outcome, should also shed new light on  why some policies succeed faster than expected, while other policies fail.

\section*{Results}
To address these modeling questions, we compiled from several sources a unique data set consisting of 170 policies that originated at the state level, along with various policy covariates, which we normalized across sources and extended for use here \cite{tribouarticle, walker1969diffusion, boehmke2012}. Of these, 81 policies (48\%) ultimately became national policies. Not included were policies where the first state policy was passed after a national policy was passed, as this would have indicated a top-down spreading mechanism that was distinct from the bottom-up diffusion we sought to examine. We also dropped alcohol prohibition from our data due to law reversals during the diffusion process, which was inconsistent with the rest of the data. 

The data from Boehmke and Skinner \cite{boehmke2012} includes a classification of policies into 13 categories, which we carry forward in our analysis and apply to policies from other data sets. We also used Boehmke and Skinner's demarcation of 11 eras, from the Age of Reform (1820 to 1860) through the New Federalism era (1980 through present). The authors note that many of these eras correspond with higher rates of state policy adoption and hypothesize that these may be the result of states implementing new and fast-diffusing policies, or more frequent adoption of formerly avoided policies with which states were already familiar. Boehmke and Skinner also include information about the year during which each state passed a given law, which enables us to derive features from the states where each policy originated. We create categories for ideology (number of covariates $c = 3$), urbanization ($c = 1$), per capita wealth ($c = 1$), state population size ($c = 2$), and professional legislature ($c =1$). For each category, we evaluate whether the originating state for a given policy is within the top five states for that category. We include variables indicating whether any of the first five states to adopt a policy is among the top five ``initiator'' states, or whether the first state is among the top five initiator states ($c =4$), either according to colloquially ``well-known'' innovators \cite{boehmke2012} or to a quantitative definition from Boehmke  and Skinner. The ``well-known'' top five states are California, New York, Texas, Massachusetts, and Illinois. Boehmke and Skinner quantitatively define California, New Jersey, Oregon, New York, and Connecticut as the top five. Finally, we include the broad census regions to group together policies originating from certain areas of the country ($c =4$). More details of these variables are included in the Methods section.

\subsection*{Part 1: Predicting whether a state policy leads to national action}
\begin{figure}%[h!]
\begin{center}
\includegraphics[width=0.48\textwidth, trim={0 7cm 0 7.5cm}, clip]{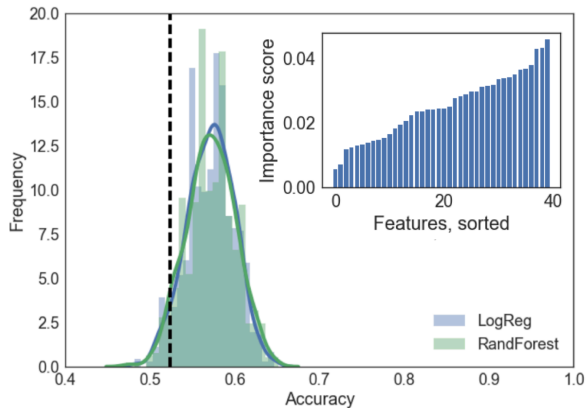}
\caption{\textbf{Accuracy on held-out data of logistic regression and random forest models for predicting whether each state-level policy yielded a national action.}
      Accuracy is computed across 1000 independent trials for each model, under 4-fold cross validation with random splitting for each trial. Accuracy is defined as the sum of true positives and true negatives, divided by the total number of policies in the data set. Both models yield approximately the same distribution of accuracy scores, which are only marginally better, on average, than the baseline classifier that guesses ``yes'' for every policy (vertical dashed line). The inset shows the importance scores for each of the 40 variables in the random forest model, averaged over 1000 trials, showing no obvious breakpoint between variables with lower vs. higher importance scores.}
\label{fig:fig1}
\vspace{-20pt}
\end{center}
\end{figure}

For each of the 170 policies in our data set, we trained two prediction models on 40 binary covariates to predict whether each of the 170 policies would result in a national action ($N=81$) or not ($N=89$). We measured the predictive accuracy of the models by counting the fraction of correct assignments for this binary metric, i.e., the number of correctly predicted true positives and true negatives, divided by the total number of policies. A simple logistic regression model using 4-fold cross validation yields an average accuracy of 0.573 across 1000 trials, or just slightly better than a trivial guess that every policy did not pass at the national level ($n_0 = 89 / 170 = 52.3$\%). The range of model accuracies ranged from 0.529 at the 5\% percentile to 0.617 at the 95\% percentile (Fig.~\ref{fig:fig1}A, blue bars).

% ------ FIGURE 1: -----
% -------------------------------------------------------

A standard random forest model, which can capture some nonlinear relationships,  using 4-fold cross validation yields an average accuracy of 0.569, which is nearly indistinguishable from the performance of logistic regression. The 5\% -- 95\% percentiles for the random forest model ranged from 0.523 to 0.612 (Fig.~\ref{fig:fig1}, green bars).

The low level of predictive accuracy shown by both models indicates that there is relatively little predictive information in the measured covariates, and the national action outcome does not strongly correlate with either linear or non-linear combinations of these features. This result indicates that many of the conclusions in the literature, as to which policy characteristics drive the spread of a policy, are not generalizable. In other words, factors relating to the importance of geography \cite{mooney2001modeling}, urbanization \cite{walker1969diffusion, shipan2012policy}, professional legislatures \cite{shipan2006bottom}, and per capita GDP \cite{walker1969diffusion, jansa2018copy}, as well as factors relating to the different categories of laws and historical time period do not in general predict which state policies become national policy. While these covariates may be important for evaluating whether an individual state will adopt a policy, they fail to produce more than a weak signal with regard to whether a policy ultimately becomes national. 

\begin{figure}%[h!]
\begin{center}
\includegraphics[width=0.48\textwidth, trim={0 6cm 0 6cm}, clip]{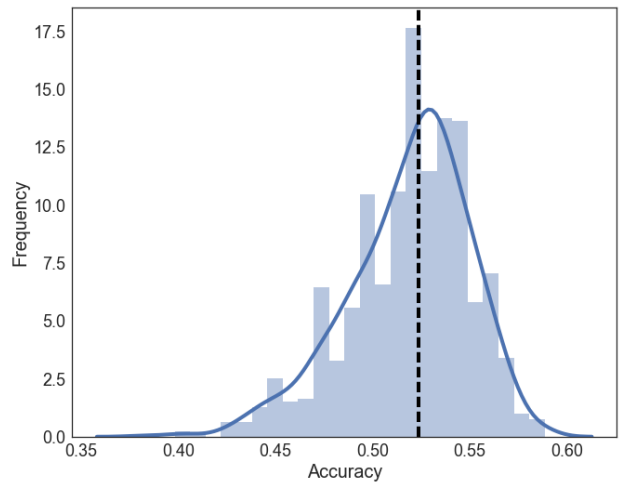}
\caption{\textbf{Accuracy on held-out data of a random forest model built using only the five features with the highest importance scores in the full model. }
      The bars are a histogram of accuracy scores over 1000 trials for a random forest model, and the blue line is the kernel density estimate. The average accuracy using these five features -- Corrections, Health, first state law passed 1980-present, first state law passed in census region South, and first five states include a Boehmke Quantitative Initiator state -- was 0.518, which is not significantly better than chance.}
\label{fig:fig2}
\vspace{-10pt}
\end{center}
\end{figure}

To determine which covariates produce this weak signal, we calculated the importance of each feature for predicting whether a policy would become a national law or not. Feature importance \cite{friedman2001elements} (also called the Gini importance or mean decrease in impurity) is a measure of how often a given feature appears near the root node of the random forest, and hence classifies a higher proportion of the data. Mathematically, we calculate a feature's importance by computing its average improvement as a splitting, where the improvement is defined as the decrease in squared error that results from using the variable to split a given tree vs. not splitting at all \cite{scikit-learn}. By this measure, the five most important features ,which appeared most frequently in the top five ranked positions were: Corrections category, Health category, first state law passed in 1980-present, state law first passed in census region of South, and Boehmke Initiator state (CA, NJ, OR, NY, CT) being one of the first five states to pass the law. However, the calculated importance scores (Fig.~\ref{fig:fig1} inset) vary smoothly, indicating that the top five features capture only a small fraction of the total importance of all features. While their frequency in the top five set was relatively high compared to the significance threshold (see Fig. ~\ref{fig:fig13S1}, supplemental materials), a model trained on only these features (Fig.~\ref{fig:fig2}), yields a mean accuracy score (over 1000 trials with cross-validation) of 0.518, which is no better than chance. The 5\% cutoff for this 5-feature model is 0.459, and the 95\% cutoff is 0.564.

% ------ FIGURE 2:  -----
% -------------------------------------------------------

Examining the data more closely indicates that health care policies post-1980 may be driving a substantial portion of the estimated importance scores. Among this set of policies, there were five that ultimately passed into national law as part of the Affordable Care Act in 2010, and an additional four policies were included in the HIPAA law of 1996. Thus, of the 14 health-related policies in the period 1980-present, nine became national law and five did not. Separately, we note that the coefficient for the covariate denoting a law originating in the South census region was negative, meaning that policies originating from this region were relatively less likely to become national laws. This finding is directionally consistent with previous research \cite{savage1978policy}.

\subsection*{Part 2: Predicting the threshold number of states for national action}
The second prediction task is to estimate the number of states that must adopt a policy before a national action makes it a national policy. A precise estimate of this number for a given policy and its covariates would provide a narrower range of estimated years in the final forecasting step for national action. Two simple hypotheses for such thresholds are that (i) the threshold varies according to the  type of policy, such that some policy covariates correlate with a lower threshold for a national action than others \cite{ nicholson2009politics}, and (ii) the threshold varies by the historical period in which the policy diffuses \cite{boushey2010policy}.

\begin{figure*}%[h!]
\begin{center}
\includegraphics[width=\textwidth, trim={0cm 10cm 1cm 10.2cm}, clip]{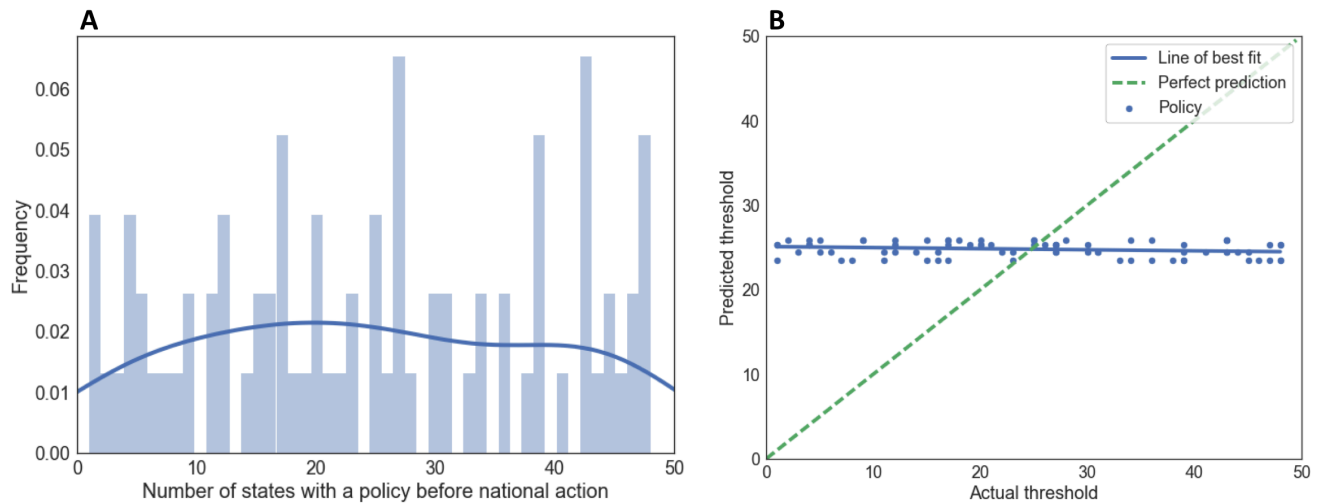}
\caption{\textbf{Predicting the threshold for national action.} (A) The number of states that have passed a state-level policy by the time a national action occurs. A kernel density estimator is shown as the blue line, illustrating the nearly uniform distribution of historical thresholds. (B) Predicted threshold as a function of the actual (historical) threshold for all policies, showing almost no correlation. Instead, the predicted thresholds simply cluster around the mean of the distribution. A simple linear fit to the predicted thresholds (blue line) has a slight negative slope, while a perfect model would follow the y=x line (green).}
\label{fig:fig3}
\vspace{-20pt}
\end{center}
\end{figure*}

For the 81 state policies that led to national action, the distribution of thresholds is nearly uniform on the interval $[1,50)$. Although the mean number of states is 24.8, which agrees qualitatively with the rule of thumb of 50--60\% of states \cite{tribouarticle, Warnerarticle}, this agreement is misleading as nearly as many policies had a threshold below 13 or above 37 as between these values (Fig.~\ref{fig:fig3}A).

% ------ FIGURE 3:  -----
% -------------------------------------------------------

Furthermore, there is close to no predictive information contained in the observed covariates for estimating the eventual national action threshold. Using the same 40 covariates from Part 1 and 4-fold cross-validation, we used a simple linear regression model with lasso regularization to learn to predict the national action threshold. For comparison, we also trained a random forest model, testing the correct max-depth parameter as a proxy for regularization. 

Plotting the predicted threshold for each policy as a function of the true historical threshold, we find no correlation under the linear model (Fig.~\ref{fig:fig3}B), even with an optimized lasso penalty (also called the tuning parameter) of $\lambda = 0.056$ \cite{friedman2001elements}. Other choices produce models with larger feature sets, but no better predictive accuracy. Random forests produced similar results, with the predictive accuracy occurring with a max-depth parameter of 1, indicating that the smallest feature set performed best.

\begin{figure}%[h!]
\begin{center}
\includegraphics[width=0.48\textwidth, trim={0 8cm 0 7cm},clip]{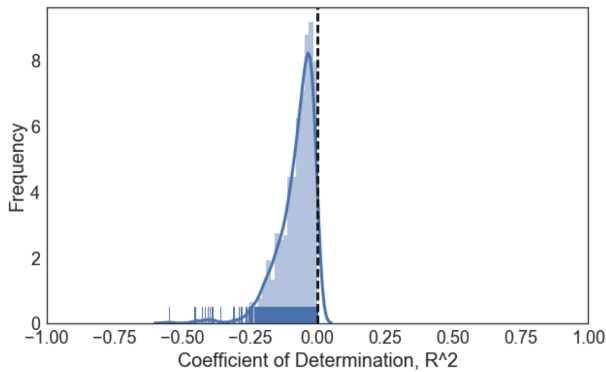}
\caption{\textbf{Histogram of coefficient of determination $R^2$ values from 1000 trials using linear model with lasso regularization.} The light blue area is a histogram showing the frequency of $R^2$ values, along with a kernel density estimate of the distribution. Vertical lines along the x-axis show the underlying $R^2$ values, all of which are bounded above by the null model of $R^2=0$ (black dashed line).}
\label{fig:fig4}
\vspace{-20pt}
\end{center}
\end{figure}

In fact, the trivial model of simply predicting the mean empirical threshold for every policy, independent of any policy covariates, performed better than any linear model. Over 1000 randomized training/test splits, the distribution of the values for the coefficient of determination $R^2$ (\cite{cameron1997r}) is upper-bounded by ${R^2} = 0$ (Fig.~\ref{fig:fig4}), indicating that a threshold equal to approximately 25 states out of 50 is a better predictor of held out data and that, even with regularization, the more flexible models are overfitted. (Additional detail on the computation of the coefficient of determination $R^2$, and how it becomes negative, are given in the Methods section.)

% ------ FIGURE 4:  -----
% -------------------------------------------------------

The nearly uniform distribution of historical thresholds is itself a notable pattern, but the lack of correlation between threshold and policy covariates indicates that these covariates, many of which are grounded in past investigations of policy diffusion, are largely meaningless when it comes to predicting the general features that presage national action. One interpretation of this negative result is that the precise circumstances and characteristics of a state-level policy becoming a national policy are unique to each policy, and there are no generalizable rules or patterns that allow ex ante prediction at rates better than chance. 

\subsection*{Part 3: Forecasting the year of national action}

The final prediction task is to estimate the particular year in which a state-level policy will become national-level policy, given only information on the timing of the first few state-level adoptions. That is, we seek to predict how quickly a policy will diffuse across the states to become a national policy. We begin by modeling two policies that recently became national law: same sex marriage (Supreme Court decision in 2015) and methamphetamine (henceforth ``meth'') precursor control laws (Congressional legislation in 2005).  These two cases allow us to quantitatively evaluate the accuracy of a simple diffusion-based forecasting model when trained on the timing data of the first five or the first 10 states that passed the law at the state level.

Because our predictive modeling results so far indicate that (i) whether a state policy eventually becomes national policy and (ii) the threshold at which it would, if it did, are largely unpredictable given available data, we model the diffusion of a policy as a simple logistic growth function for the number of states that have adopted the policy up to some time $t$. The shape of this curve is fully determined by the date of the initial state adoption and a growth rate parameter. Given an estimated model and then a choice of threshold for how many states must adopt the policy before a national action, we derive a precise year at which the logistic function crosses that threshold. 

For each policy considered, in each repetition, we choose a threshold uniformly at random from the historical thresholds from our dataset of  national policies. This approach allows us to capture, in a data-driven way, the induced uncertainty from the unpredictability of the thresholds in our forecasts.

To better capture the underlying variability in the timing in which each state adopted a policy, we employ a smooth bootstrap and we fit the logistic growth model independently to each such bootstrap of the data. To generate a smooth bootstrap sample, we first choose a standard bootstrap of the observed years, and then add normally distributed noise, with zero mean and unit variance.

As a result, each bootstrap sample is a set of continuous values. The logistic growth parameter is then estimated from the bootstrap sample by minimizing the sum of squared errors. The output of this procedure is a distribution of the year at which the logistic function crosses the national action threshold, which we interpret as a kind of posterior distribution for when a policy will become a national law.

\begin{figure*}%[h!]
\begin{center}
\includegraphics[width=\textwidth, trim={0cm 9.5cm 0cm 9.8cm}, clip]{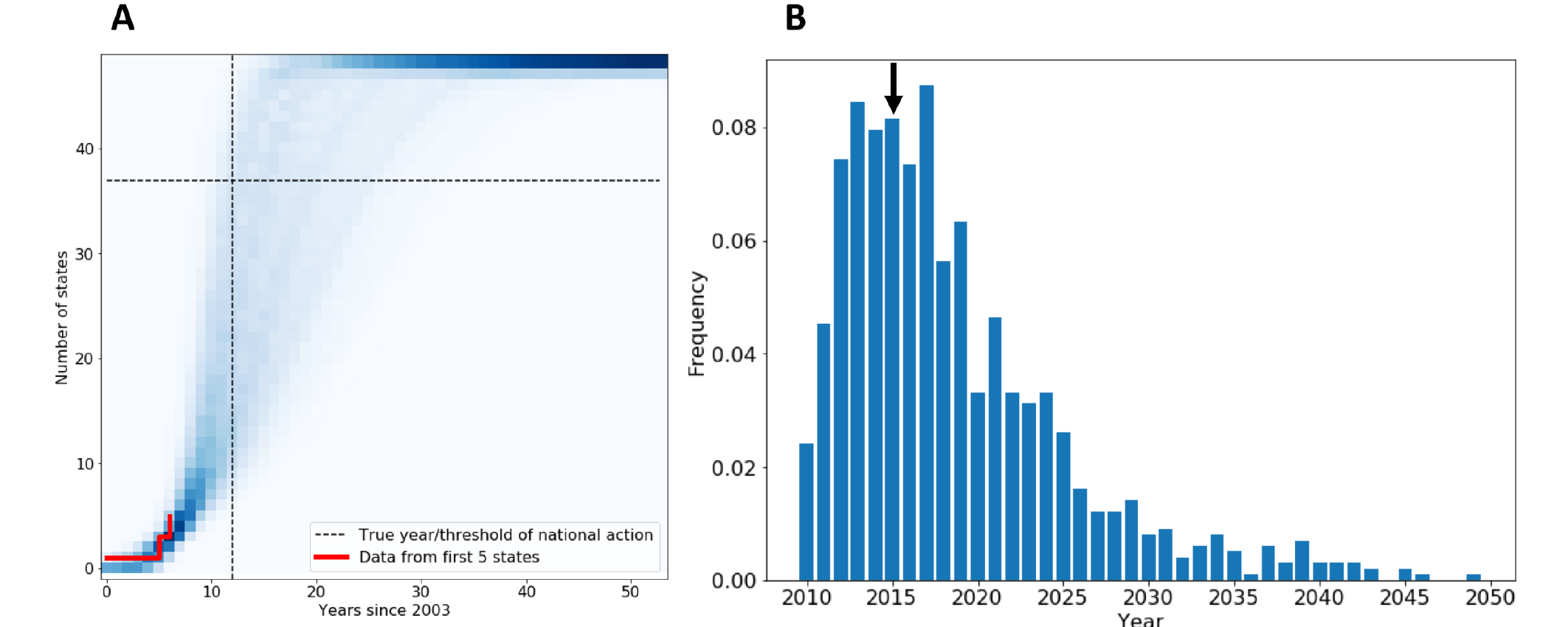}
\caption{\textbf{Forecast for same sex marriage using only data from the first five states. } (A) A heatmap showing the posterior density of policy diffusion trajectories, i.e., the cumulative number of states having adopted a policy as a function of time, under our model (see text). The historical trajectory for the first five states is shown in red, and the historical threshold and year of national action are shown as black dashed lines. (B) The distribution of predicted years of national action using a bootstrapped threshold. The black arrow indicates the actual year that national action was taken, 2015.}
\label{fig:fig5}
\vspace{-10pt}
\end{center}
\end{figure*}

\begin{figure*}%[h!]
\begin{center}
\includegraphics[width=\textwidth, trim={0cm 9.5cm 0 9.5cm},clip]{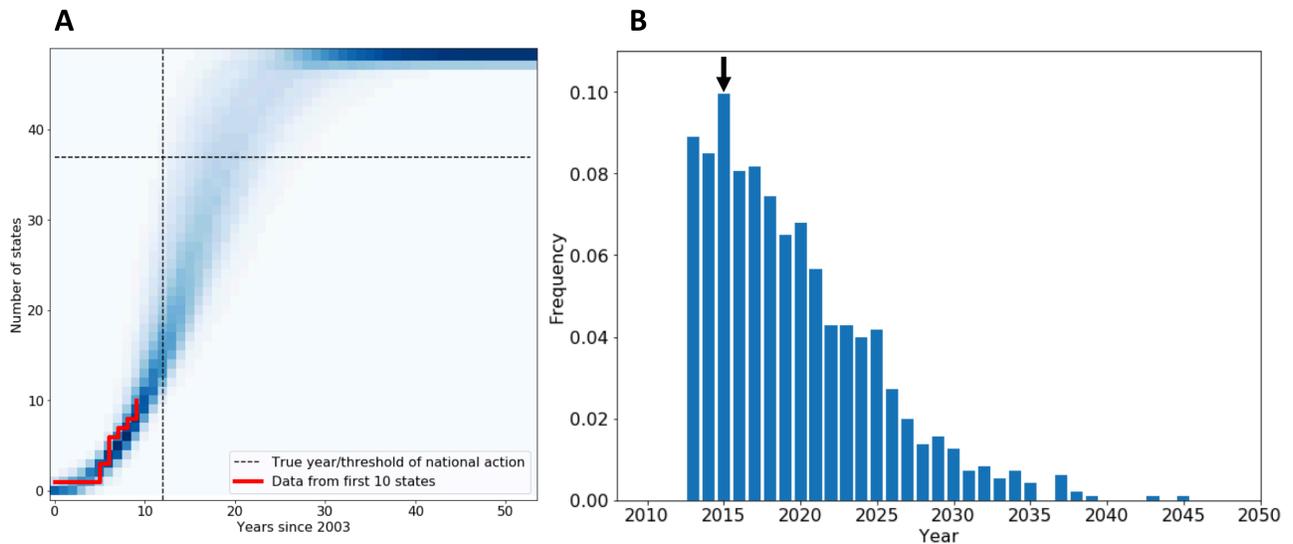}
\caption{\textbf{Forecast for same sex marriage using the first ten states to pass a state-level same sex marriage law.} (A) A heat map of the projected trajectories from our model (see text) based on the first 10 states to pass same sex marriage, shown by the red line. The historical threshold and year of national action are shown as black dashed lines. (B) The predicted years of national action, based on the logistic growth model and bootstrapped thresholds. The black arrow indicates the actual year of national action, 2015.}
\label{fig:fig6}
\vspace{-20pt}
\end{center}
\end{figure*}

Same sex marriage provides a useful test case for the diffusion model because of its relatively fast spread: 11 years from first adoption in 2004 by Massachusetts to a 2015 national action by the Supreme Court in deciding Obergefell vs. Hodges.  For this policy, the diffusion model is remarkably accurate: when trained on the first five years of data, 2015 is the among the three years with the highest probability for national action in the model (Fig.~\ref{fig:fig5}B), while it is the most likely year when trained on 10 years of data (Fig.~\ref{fig:fig6}B).

% ------ FIGURE 5:  -----
% -------------------------------------------------------

% ------ FIGURE 6:  -----
% -------------------------------------------------------

The accuracy of these estimates conceals some uncertainty in the form of a broad distribution centered around these values. Training on either the first five states or 10 states concentrates about 8\% or 10\% of the density on a 2015 date for national action, respectively. However, the models place 38.7\% and 43.6\% respectively in the forecast range of $2015 \pm 2$ years, indicating a strong concentration of density around this year, with only 18.4\% and 17.5\% of density falling 10 years or more after 2015. As a small note, training on 10 versus five years of data does increase the concentration of forecast density around the mode, mainly by eliminating slower diffusion trajectories (see Fig.~\ref{fig:fig5}A vs. Fig.~\ref{fig:fig6}A).

Meth precursor laws provide another useful test of the simple diffusion model. These laws control the sale and distribution of over-the-counter drugs that contain pseudoephedrine (e.g., Sudafed), which is used in the production of meth. The first such state law was passed by Oklahoma in 1996, and a national law was passed by Congress in March 2006 after a total of 25 state laws had been passed.

\begin{figure*}%[h!]
\begin{center}
\includegraphics[width=\textwidth, trim={0cm 9.9cm 0 9.9cm},clip]{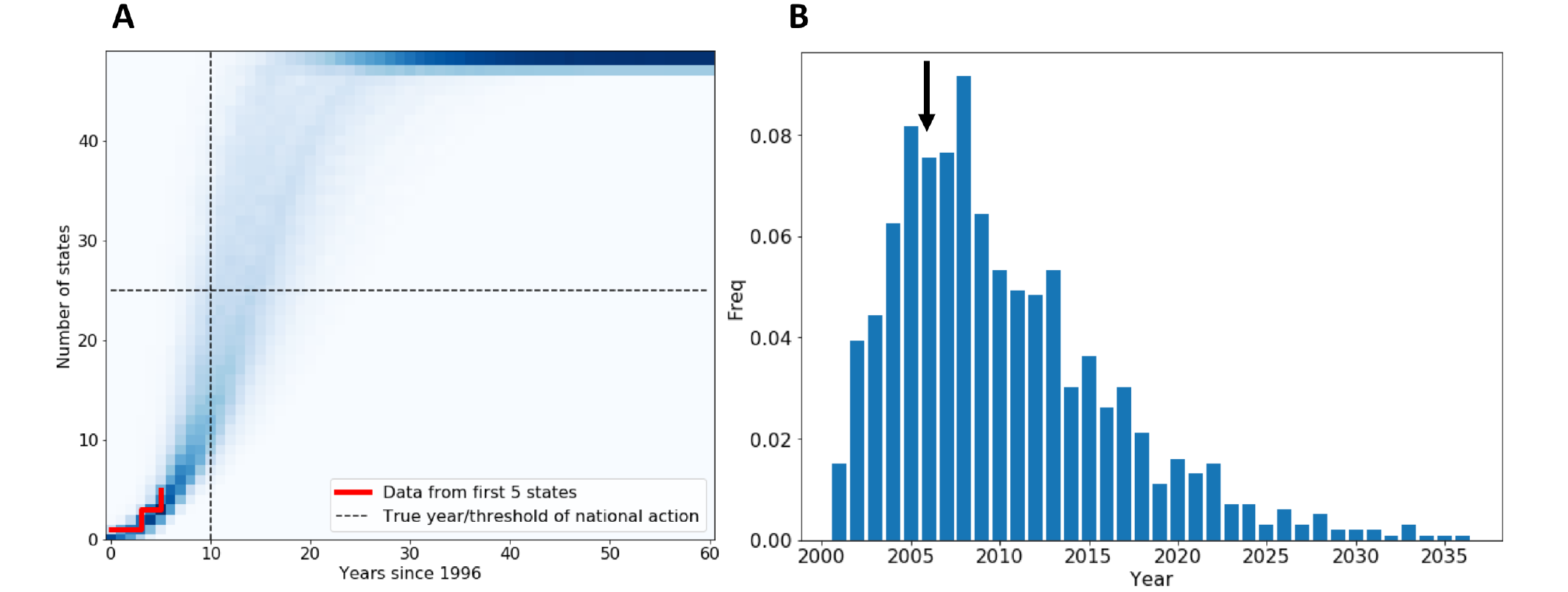}
\caption{\textbf{Forecast for meth precursor control laws using the data from the first five states.}    (A) The historical trajectory for the first five states is shown in red, and the heat map shows the posterior density of trajectories from the diffusion model. The dotted black lines indicate the historical year of national action and the historical threshold. (B) The predicted year of national action, based on the logistic growth model and bootstrapped thresholds. The black arrow indicates the year when a national meth control law was passed by Congress, in 2006.}
\label{fig:fig7}
\vspace{-20pt}
\end{center}
\end{figure*}

\begin{figure*}%[h!]
\begin{center}
\includegraphics[width=\textwidth, trim={0cm 9.7cm 0 9.5cm},clip]{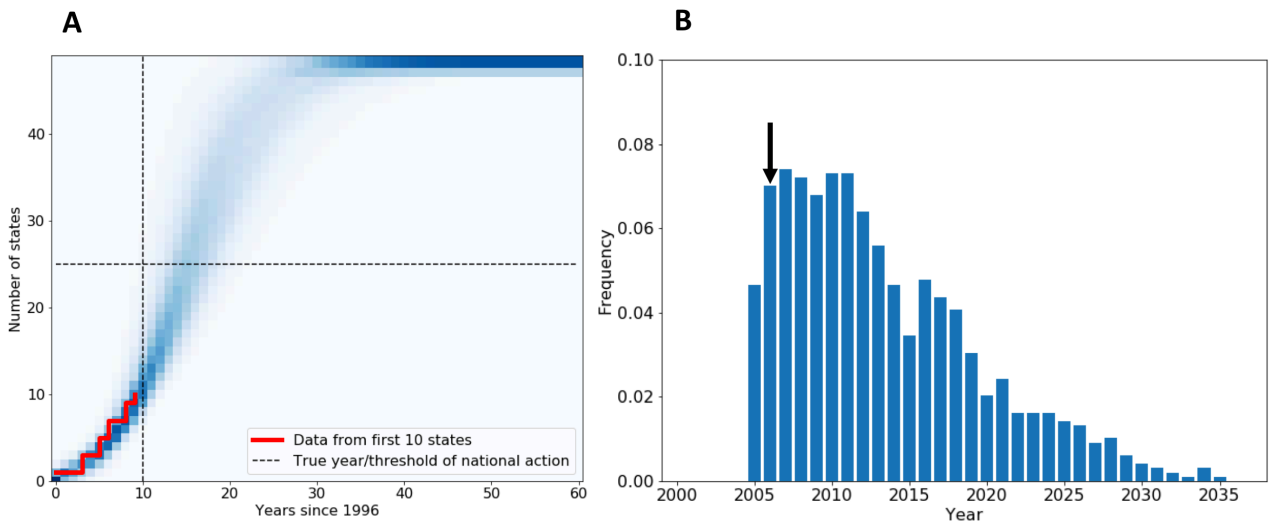}
\caption{\textbf{Forecast for meth precursor laws using the first ten states.} (A) The red line shows the historical trajectory for the first ten states to pass this law, and the heat map shows the posterior density of trajectories from the diffusion model. The dotted black lines show the historical threshold and year of the national law. The red line nearly intersects the dotted black line because 15 additional states passed a meth precursor law in late 2005, before a national law was passed in early 2006. (B) The predicted year of national action, with the historical year of national action in 2006 indicated by the black arrow.}
\label{fig:fig8}
\vspace{-20pt}
\end{center}
\end{figure*}

Figures~\ref{fig:fig7} and~\ref{fig:fig8} show the posterior distributions of adoption trajectories for meth precursor laws, when trained on the timing of the first five (Fig.~\ref{fig:fig7}A) and first 10 states (Fig.~\ref{fig:fig8}A) to pass such a law. What is hidden from these estimates is a very large group of simultaneous adoptions in the latter part of 2005, when 15 states passed meth precursor laws, effectively shifting the trajectory to a much steeper path. Nonetheless, the bootstrapped thresholds control for this uncertainty to some degree and the distribution of predicted years of national action remains concentrated around the historical threshold. The exact year of 2006 only contains 7.6\% of the density for national action using five years, and 7.0\% using 10 years of data. However, both predictions place the probability of national action at 38.8\% (Fig.~\ref{fig:fig7}B) and 26.3\% (Fig.~\ref{fig:fig8}B) respectively when looking at $2006 \pm 2$ years, with the 5-year prediction being more accurate with trajectories centered around the actual year of national action. Only 17.7\% and 32.2\% of the probability density falls 10 years or more after 2006.

% ------ FIGURE 7:  -----
% -------------------------------------------------------

% ------ FIGURE 8:  -----
% -------------------------------------------------------

For this policy, data on the timing of the first five states led to better forecasts than data on the first 10 states, likely because states laws in states numbered 11 through 25 passed meth precursor laws in 2005 \cite{mcbride2008relationship}, one year before national action. As a result, the relatively slower growth rate represented by when states 6-10 passed this policy was a poorer indication of the coming burst of activity. Hence the accuracy of such a simple diffusion model's forecast, which omits parameters that might capture complex dependencies in the spread of a policy, is depends strongly on how well the empirically observed spreading dynamic follows the classic logistic growth. 

We now use this same model to make forecasts for two policies that have not (yet) had national actions. This exercise allows us to discuss some of the subtleties of the forecasting model without the benefit of hindsight. In particular, we make a data-driven forecast for the legalization of recreational marijuana and for the passage of ``stand your ground'' laws using the timing of the first five state adoptions and of all available data to date.

\begin{figure*}%[h!]
\begin{center}
\includegraphics[width=\textwidth, trim={0cm 9cm 0 9.5cm},clip]{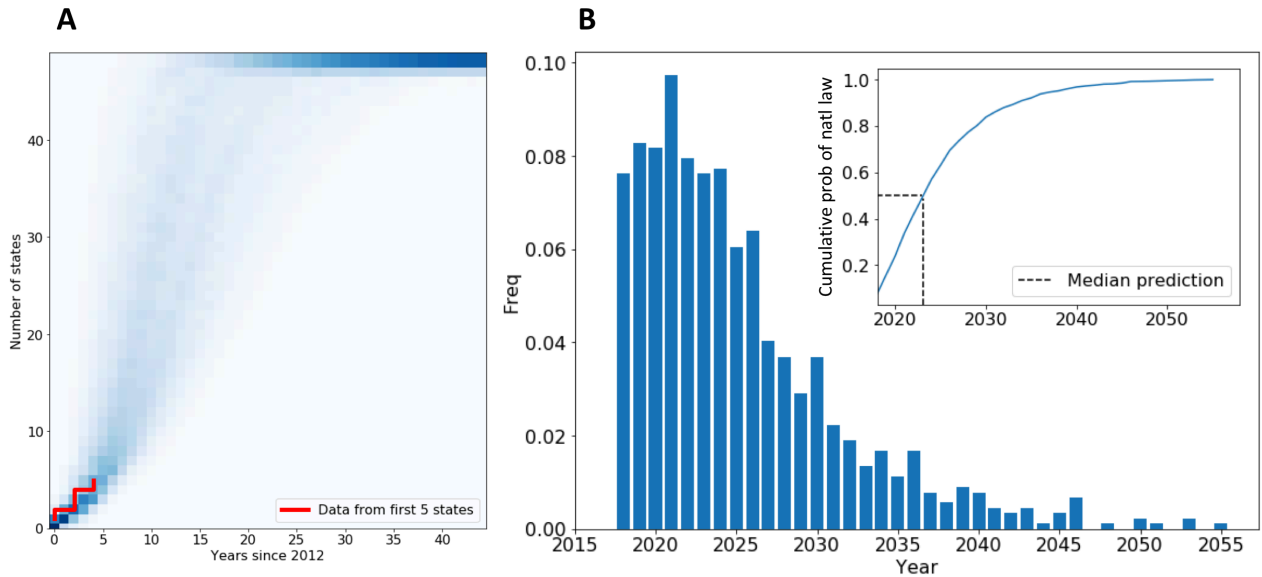}
\caption{\textbf{Forecast for national recreational marijuana legalization using the first five states.} . (A) The historical trajectory of the first five states to legalize recreational marijuana (red line), which spaces 2012--2016, along with the heatmap showing the posterior distribution of model-based trajectories. (B) Histogram of forecasted year of national action, in which 2021 is the maximum a posteriori estimate of individual year of national action. The cumulative probability distribution in the inset figure shows that the median year for national action, where the cumulative probability crosses the 50\% line, is 2023.}
\label{fig:fig9}
%\vspace{-20pt}
\end{center}
\end{figure*}

\begin{figure*}%[h!]
\begin{center}
\includegraphics[width=\textwidth, trim={0cm 9.2cm 0 9.5cm},clip]{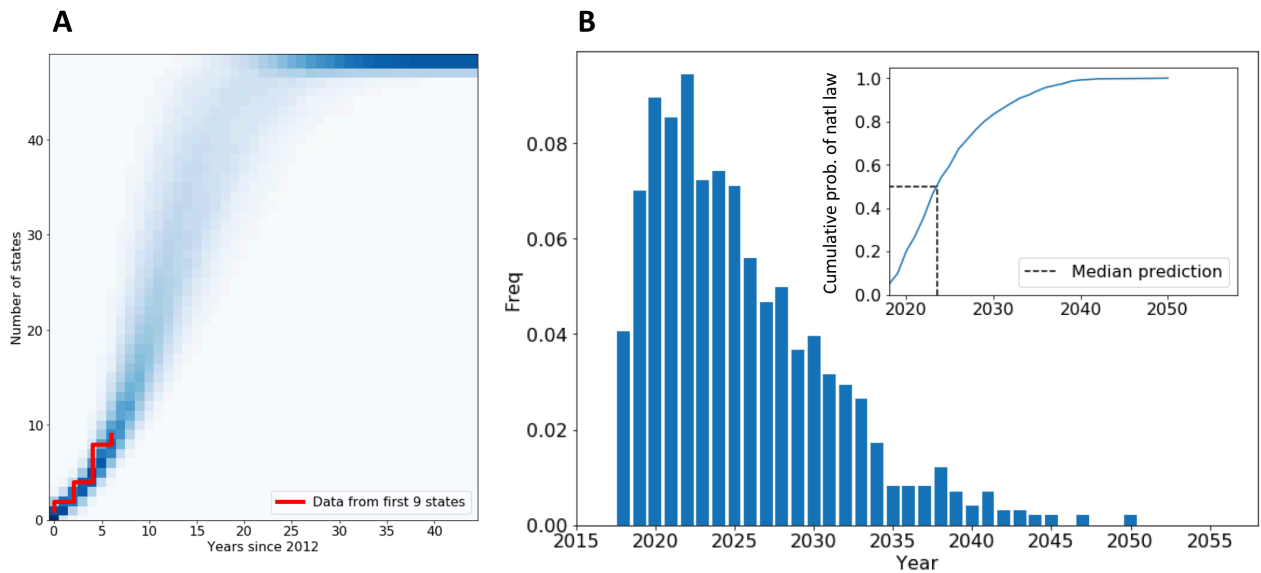}
\caption{\textbf{Forecast for national recreational marijuana legalization using the first nine states. }  (A) The historical trajectory of these state adoptions is indicated by the red line. The heat map shows the posterior density of predictive trajectories. (B) Histogram of the forecasted year for national recreational marijuana legalization. The cumulative probability distribution in the inset figure shows that the median year for national action, where the cumulative probability crosses the 50\% line, is 2023.}
\label{fig:fig10}
\vspace{-10pt}
\end{center}
\end{figure*}

Recreational marijuana laws were first adopted in 2012 and by July 2018 had been adopted by 9 states. We make a forecast for the year of national action using the timing of the first five states (Fig.~\ref{fig:fig9}) and then subsequently using the timing of all nine states (Fig.~\ref{fig:fig10}).

% ------ FIGURE 9:  -----
% -------------------------------------------------------

% ------ FIGURE 10:  -----
% -------------------------------------------------------

Training on either the first five states or nine states concentrates the density of the most likely years for national action: about 9.7\% on 2021 (Fig.~\ref{fig:fig9}B) or 10.5\% on 2023 (Fig.~\ref{fig:fig10}B), respectively. However, the models place 42\% and 35.7\% respectively in the forecast range of the next five years, by the end of 2022, with only 26\% and 28.2\% of density falling after 2028. The 9-state prediction (Fig.~\ref{fig:fig10}A) eliminates some of the steeper trajectories seen in the 5-state prediction (Fig.~\ref{fig:fig9}A), lowering the probability of national action in 2018 or 2019, although the median of the posterior distribution remains at 2023 in both models (Fig.~\ref{fig:fig9}B inset, Fig.~\ref{fig:fig10}B inset).

\begin{figure*}%[h!]
\begin{center}
\includegraphics[width=\textwidth, trim={0cm 9.8cm 0 9.5cm},clip]{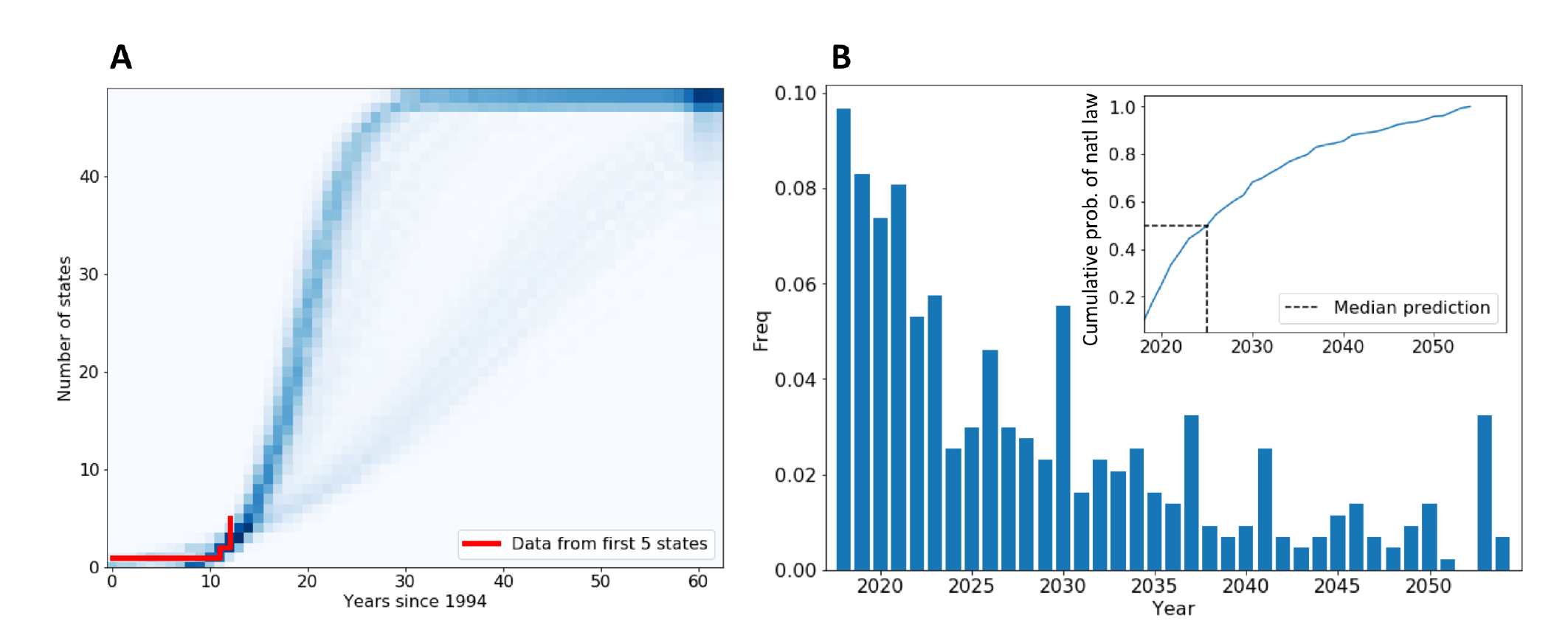}
\caption{\textbf{Forecast for ``stand your ground'' laws using the first five states.} (A) Heatmap of the projected trajectories for ``stand your ground'' laws, based on historical data from the first five state-level laws (red line). The first state-level law was passed in 1994, and the next state passed its law in 2005. (B) The posterior distribution of forecasted year of national action on ``stand your ground'' laws. The inset shows the cumulative probability distribution with a median predicted year of 2025.}
\label{fig:fig11}
\vspace{-20pt}
\end{center}
\end{figure*}

``Stand your ground'' laws had very slow adoption after the first such state law passed in 1994, until a large pulse of state laws passed simultaneously in 2006 (Fig.~\ref{fig:fig11}A). As a result, the diffusion of this policy deviates substantially from the smoother trajectory expected by our simple model. Nevertheless, our statistical forecast provides a data-driven view of its likely future diffusion. 

% ------ FIGURE 11:  -----
% -------------------------------------------------------

% ------ FIGURE 12:  -----
\begin{figure*}%[h!]
\begin{center}
\includegraphics[width=\textwidth, trim={0cm 10cm 0 10cm},clip]{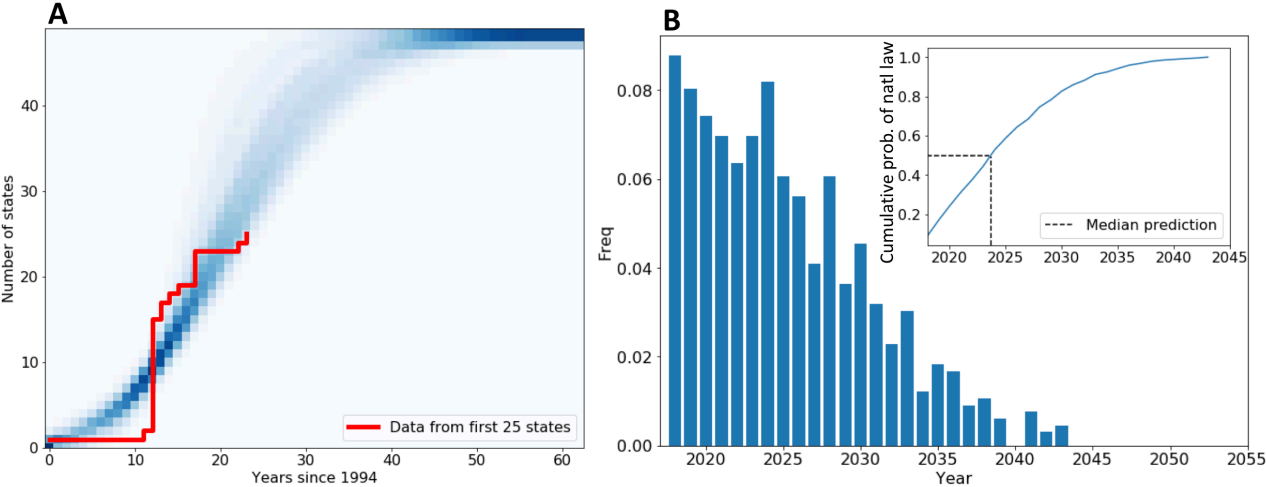}
\caption{\textbf{Forecast for ``stand your ground'' laws using the first twenty-five states. } (A) Heatmap of the projected trajectories for this policy, based on historical adoptions by 25 states (red line). (B) The posterior distribution of forecasted year of national action, under our model (see text). The inset shows the cumulative probability distribution with a median predicted year of 2023.}
\label{fig:fig12}
\vspace{-20pt}
\end{center}
\end{figure*}
% -------------------------------------------------------

Unlike the recreational marijuana policy, this policy already has 25 state laws. Thus, the ensemble of diffusion models produces a generally decreasing function for the forecasted year of national action (Figs.~\ref{fig:fig11}B and~\ref{fig:fig12}B), because most bootstraps choose a threshold that has already been surpassed. The median year for the cumulative probability function is 2025 when trained on five years of data (Fig.~\ref{fig:fig11}B inset), and 2023 when trained on all 25 years (Fig.~\ref{fig:fig12}B inset). Both models predict national action is most likely in 2018, with density of 9.8\% and 11.0\% respectively. The models place 38.7\% and 41.8\% respectively in the forecast range of 2018 through 2022, with 42.4\% and 30.9\% of density falling 10 years or more after 2018. The model using all 25 years (Fig.~\ref{fig:fig12}A) eliminates a set of trajectories from the 5-year model that project a national action occurring after 2050 (Fig.~\ref{fig:fig11}A). These trajectories place more density in the year 2054 than in the year 2024, reflecting an overfitting of the slow part of the initial trajectory where only one state had passed a ``stand your ground'' law between 1994 and 2005 (Fig.~\ref{fig:fig11}B). The majority of the trajectories do capture the uptick after 2005, but this artifact causes a broad range of forecasted years. As a result, the posterior probability of the law being passed more than ten years out is higher than any other law we examined, with 42.4\% of the density in Fig.~\ref{fig:fig12}B.  When we incorporate all of the available data from 25 states, estimates beyond 2043 are eliminated as are the visual shallow-sloped trajectories that were seen in Fig.~\ref{fig:fig11}A.

\section*{Discussion}
Across the three predictions tasks, we find widely different results in the degree to which policy diffusion outcomes can be predicted from data. The first two tasks focused on covariate-based predictive models, in which we either (i) aim to predict the binary outcome of whether a state-level policy will eventually yield a national action or (ii) aim to predict the threshold number of state-level adoptions prior to that national action. In both cases, we find that the policy covariates are extremely poor predictors of the outcome variables. These results indicate that much remains unknown about the true determinants, if any, of successful policy diffusion.
 
A number of these covariates represent existing hypotheses in the policy diffusion literature, e.g., national actions are more likely, and at lower thresholds, for policies that begin spreading from innovative states or that focus on topics like child safety. However, the lack of predictive power of these covariates in our analysis indicates that these explanations do not generalize. That is, while they may be accurate explanations for some specific policies, or under special circumstances, they are not generally predictive.

This negative result is consistent with the radical hypothesis that there is no general predictive power in any policy covariate, and that the spread of every policy is fully contingent on unique circumstances that describe the particular political circumstances in which that policy is formulated and advanced. This extreme possibility, which we call the 'unique circumstances hypothesis,' is a useful null expectation against which future work on policy diffusion can push. That is, we suggest that in the future, the default assumption should be that no policy features are expected to predict the two aspects of policy diffusion investigated here, unless shown otherwise through a systematic evaluation. Of course, the covariates in our analysis are a subset of possible features that describe the space of policies, and there may exist some features that do provide non-trivial predictive power of policy spread and eventual national action. A valuable line of future work would be to identify and derive additional features and test their predictive power in a similar setting as considered here.

We recognize that past research did find evidence that certain factors are important for policy diffusion. However, our meta-analysis of a large set of policies did not identify any predictive power in those particular covariates. These differing results can be explained by previous work generalizing from a biased subset of policies. Past hypothesized factors may have had some influence on national policy, but our results show that different factors are associated with just as many failures as successes. Focusing on the successes alone, e.g., through case studies, may have given rise to a false impression of generality. For example, with the special status of states like CA or TX or MA, our results indicate that those states have indeed generated some national policies, but they have also produced a roughly equal number of policies that spread only partially and never became national, and similarly, other states are just as likely to originate both types of policies. 

In contrast, the simple diffusion model, based on classic logistic growth, performed very well at forecasting the diffusion dynamics, providing accurate distributions of dates for national action on two known policies (same sex marriage and meth precursor laws) and plausible forecasts for two currently spreading policies (recreational marijuana legalization and ``stand your ground'' laws). We look forward, in particular, to observing how our forecasts for the latter policies fare in the future.

Notably, this simple diffusion model contains no information about the identity or character of individual states, their relationship with each other (geographic, economic, cultural, etc.), or features related to their legislative activities. Hence, a valuable line of future work would take the diffusion model as a baseline and incorporate the effect of policy covariates on the timing dynamics in a model-driven way. This more flexible model could then serve as a useful tool for making more fine-grained forecasts about different types of policies and provide a tool for understanding how policy characteristics influence the spreading dynamics.

Another notable omission in the simple diffusion model is the lack of both the political context, such as the support or opposition from the executive branch, and the level of public support for a policy. For instance, public support for same sex marriage when the policy first began spreading in 2004 \cite{pew2015changing} was substantially lower than was public support for recreational marijuana legalization in 2012 \cite{newport2012americans}. As a result, the diffusion rate for recreational marijuana legalization may be substantially higher than it was for same sex marriage. Exploring the effect of these contextual variables on the diffusion of policy, within a predictive-modeling framework, would be a useful direction of future work.

One key feature of the modeling approach undertaken to forecast the diffusion of a policy was the incorporation of uncertainty. This feature was accomplished in two ways: first, via a bootstrap of the historical timing dates of state-level adoptions, and second, via a randomized choice of the threshold required for a national action. Incorporating these stochastic elements produces an ensemble of predictive models that naturally capture some amount of underlying variability in the complicated political processes that produce the precise dates of state adoptions. The resulting distributions of predictions can then be interpreted as a probabilistic forecast, rather than as a single prediction. The use of such probabilistic approaches in political forecasting is crucial, as simple models will rarely capture all the causal factors that shape these complex social processes. The tradeoff, however, is a less certain prediction. In our view, this tradeoff is essential, and highlights the provisional nature of the output of such models: they are most useful for gaining insight and testing scientific hypotheses, and not for making policy themselves.

The diffusion of policies through the states is a crucial phenomenon in the federalist system of governance in the United States. However, the mechanisms that drive the spread of policies are poorly understood, and most existing analyses have focused on explaining the circumstances of individual policies and their diffusion, or on understanding one particular feature and its effect on a handful of policies. Systematic and data-driven analyses, like the one undertaken here using a predictive modeling approach, are relatively rare, but represent a promising direction for future work. Such analyses can test existing hypotheses, showing for instance that few or none of the previously suggested policy features are in fact generally predictive of national actions, and identify novel patterns that deserve theoretical explanation. We look forward to further investigations along these lines.

\begin{acknowledgments}
The authors thank Mike Weissman, Brian Keegan, Rafael Frongillo, Mike Mozer, Abigail Z.\ Jacobs, Jordan Boyd-Graber, and Chenhao Tan for helpful conversations. The authors also thank Nicole Beckage for input on the manuscript.
\end{acknowledgments}

\section*{Materials and Methods}

\subsection*{The data}

Our data on state-level policies, their characteristics, and national action outcomes, are compiled from several existing databases, and extended by several hand-coded variables of interest to this study. 

The first broad compilation of state policies was assembled for a paper published by Walker in 1969 \cite{walker1969diffusion}. The full data set was downloaded from ICPSR \cite{ICPSR}, the Inter-University Consortium for Political and Social Research, and includes data on 85 policies enacted between 1813 and 1966 along with their characteristics. Boehmke and Skinner \cite{boehmke2012} reproduced Walker's original article in 2012 using more recent policies and archived their data set on the Harvard Dataverse \cite{dataverse}. This data set includes information on 137 policies enacted between 1913 and 2009, of which 95 are distinct from the original Walker dataset, along with policy covariates related to each policy type and the era in which it was passed. 

To these, we manually added 6 policies from the Bloomberg data set \cite{tribouarticle}, which was assembled to visualize the diffusion trajectories of policies for an online article. This data set contributed information on two policies of interest: same sex marriage and recreational marijuana legalization. The information on these policies was updated by hand to include the passage of additional state recreational marijuana laws through 2018, and the passage of Obergefell vs. Hodges, legalizing same sex marriage at the national level.

Together, these data sets comprise 186 policies. From these, we selected the 170 policies with the following property: if the policy was enacted at the national level, the policy must have originated in at least one state before subsequent national adoption. Fifteen policies were excluded because the date of national action matched or preceded the date that the first state-level policy was passed, which indicates a top-down mechanism of state-level adoption, rather than the bottom-up diffusion process of interest here. Additionally, we excluded alcohol prohibition, due to the many state law reversals during its diffusion process, which are inconsistent with the rest of the data.

A key advantage of the data sets combined here is that slightly different versions of a given policy, passed by different states, have been aligned and normalized by experts prior to analysis. For example, gun safety regulations targeted at children vary considerably in their details across state lines. The Boehmke, Walker, and Bloomberg data sets mitigate this issue by identifying and grouping comparable state laws based on their equivalent purpose \cite{walker1969diffusion, boehmke2012, tribouarticle}. The compiled and extended data set is included in the Supplementary Materials.

Finally, we exclude the District of Columbia from all analyses and ignore the years in which states were admitted to the Union. Some studies of policy diffusion have specifically incorporated the maximum numbers of states that could adopt a policy at a given time, based on the total number of states in the Union. However, most policies that do lead to a national action do this long before approaching the upper limit of the extant number of states (Fig.~\ref{fig:fig3}A). To simplify our analysis, we thus treat states as all having existed throughout the time period we consider. For example, 21 of the 170 policies in our data set were first passed at the state level before 1892, when there were no more than 44 states. 

\subsection*{Policy covariates}

To create a consistent and sufficiently detailed corpus of policies, we combined existing policy covariates included in the source data sets with a set of additional hand-coded covariates. Existing policy covariates included indicator variables of the general topic of each policy across 13 categories: administration, civil rights, conservation, corrections, education, elections, health, labor, planning, professional (i.e., licensures), taxes, and welfare. Among this set, health and corrections policies were overrepresented. However, each category had a minimum of 6 representative policies. Also included were indicator variables for the historical era in which the policy was diffusing. While other authors assign a policy to the historical era (i.e. the politically and economically notable time period) in which the tenth state passed the law \cite{boehmke2012}, this rule would be inapplicable to policies that became a national law with less than 10 state adoptions. To accommodate these policies, we re-coded all era covariates to the era in which the first state-level policy was passed.

Hand-coded covariates included indicators for the census region of the first state where a policy was passed and additional statistics for that state: per capita income, urbanization, and population size. These indicator variables capture the hypothesis that some regions of the country are more innovative than others \cite{shipan2006bottom, boehmke2017seeds} or that states with professional year-round legislatures are more likely to innovate \cite{ jansa2018copy}. Professional (year-round) legislature information came from King \cite{king2000changes}. We also hand-coded variables to indicate whether the first state to pass a given policy was among the five states with the largest population, smallest population, largest per-capita wealth, degree of urbanization, most liberal, most conservative, or most extreme in either ideology \cite{ grossback2004ideology}. Population numbers were extracted from the US Census Statistical Abstract of the States \cite{uscensus} data set and the corresponding covariates coded as the average state population in the census between 1960 and 2010. Per-capita income was coded as the average of per-capita income in 2005 dollars from 1980--2010. Urbanization is a recent measure only, and so we coded this variable as the average urbanization score in the US census from 1990 and 2000 \cite{uscensus}. Finally, a state's ideology was coded as the average score for the state's ideology (as a function of state governor's party, roll call voting scores of state congressional delegations, outcomes of congressional elections, etc.) from 1960--2006, as described in \cite{ berry1998measuring}.

The final hand-coded covariates indicated whether a state policy originated in one of the top-5 ``innovative states,'' either as the first state to pass such a law, or if an innovative state was among the first five states to adopt the policy. Because definitions of innovativeness differ across the policy diffusion literature, we used two separate lists of most-innovative states. The first, defined by \cite{boehmke2012} as ``well-known'' innovators, consisted of California, New York, Illinois, Massachusetts, and Texas. The second list, based on quantitative evaluation also by \cite{boehmke2012}, consisted of California, New Jersey, Oregon, New York, and Connecticut.

\subsection*{The coefficient of determination}
To examine model quality for the threshold determination prediction task in Part 2, we used a standard definition \cite{cameron1997r} of the coefficient of determination
\begin{equation}
  R^{2} = 1 - {\rm SS}_{\rm res} / {\rm SS}_{\rm tot} 
\end{equation}
where $SS_{\rm res}$ is the sum of the residual sum of squares, and $SS_{\rm tot}$ is the total sum of squares. These in turn are defined as the following:
\begin{align}
 	 {\rm SS}_{\rm res} & =  \sum_i \left( y_{\rm true}(i) - y_{\rm pred}(i)\right)^2 \\
          {\rm SS}_{\rm tot}  & = \sum_i \left(y_{\rm true}(i) - y_{\rm mean}) \right)^2
\end{align}
Because the coefficient of determination $R^2$ is measuring the improvement of the model relative to a constant model where $y_{\rm pred} = y_{\rm mean}$, there is a possibility for negative values if the model is worse than a simple model of the mean, which was the case in Part 2 of the analysis. Notably, there are other definitions of $R^2$ that represent different mathematical concepts, such as the fraction of variance explained. These should not be confused with the coefficient of determination used here. 

\subsection*{The diffusion model}

Our simple model of policy diffusion is a classic logistic growth model, which describes how the total number of states that have adopted a policy grows over time. This model has previously been used to describe policy diffusion although primarily in a qualitative fashion as an S-curve \cite{boushey2010policy}. Our use of it here as a parametric model for curve-fitting and statistical forecasting more closely follows the use of \cite{Warnerarticle}.  A standard form for logistic growth is
\begin{equation}
  P(t) = \frac{K P_0 \textrm{e}^{rt}}{K + P_0 (\textrm{e}^{rt} - 1)}
\end{equation}
where $P$ is the current percentage of states with a given policy in place, $t$ is time since the first such policy was introduced, $K$ is the carrying capacity (in this case, $K = 50/50 = 1$) and $r$ is the growth rate of the diffusion process.  Thus, this model is fully determined by two parameter choices: $P_0$, the initial number of states that adopt a policy, and $r$, the rate of growth. In practice, $r$ is the more important parameter, as it sets the speed of the diffusion, and $P_0 = 1$ is typical. 

Although this logistic growth model is continuous in both time $t$ and population $P$, it can be applied here, where both $t$ and $P$ are rational values, without significant loss in accuracy. To do so, $P_0$ is typically set equal to $1/50$, as most policies begin in a single state. However, by allowing this parameter to vary slightly, we often obtain better fits to the empirical data, while still requiring $P_0$ to be positive.  Including this flexibility in our parameter estimation mitigates the mismatch between the continuous logistic growth model and the discrete nature of the data. The only constraint placed on $r$ is that it is positive.

These parameters are then chosen via simple grid-search optimization of the sum-of-squared errors, 
\begin{equation}
  	{\rm SS} = \sum_{t=1}^{n} \left( y(t) - P(t)\right)^2
\end{equation}
where $y(t)$ is the fraction of states that have passed a given policy in the $t$ years since the first state adopted it. Given a choice of model parameters, we may then extrapolate the logistic function to characterize its long-term diffusion trajectory for the policy and estimate the precise year in which the function exceeds any particular threshold for a national action.

\subsection*{Incorporating uncertainty via bootstrapping}

The underlying data generating process for the precise timing of when any particular state adopts a policy is highly complex, and our simple diffusion model omits nearly all such details. Hence, fitting only the historical sequence of years that states adopt a given policy may overfit or underfit the data and produce less reliable forecasts of the future trajectory of a policy. To account for these underlying sources of variability and uncertainty in a natural way, we use a smooth bootstrapping approach for the timing data, and a classic bootstrap for the national action threshold. We then fit each bootstrap data set with our logistic growth model and derive an estimate of the timing of the national action by calculating when the fitted curve crosses the bootstrapped national action threshold. This approach also helps control for missing or inaccurate state-level data.

To generate a single bootstrapped data set, we begin by treating each state policy adoption as an independent event. We sample $m$ of these events uniformly at random, from the historical sequence of events  $(1, 2, \dots, m)$, with replacement. Next, for each sampled date, we add noise drawn from a normal distribution, with mean of zero and variance of 1 year. This smoothed bootstrap models the uncertainty of the state policy-making processes and data recording, e.g. policies that may have been approved by voters in November 2007 and enacted in January 2008, and the year being recorded as 2007 or 2008 in different data sets. We repeat this bootstrapping process for each of the 1000 trials for a given policy, to generate a range of diffusion trajectories. Hence, there are two simple sources of variability that drive the distributions of threshold-crossing years which are data-driven and intended to present a reasonable model of the underlying uncertainty in the data-generating process.

% ------ FIGURE 13:  -----
\begin{figure}%[h!]
\begin{center}
\includegraphics[width=0.48\textwidth, trim={.5cm 5.9cm 0cm 5cm}, clip]{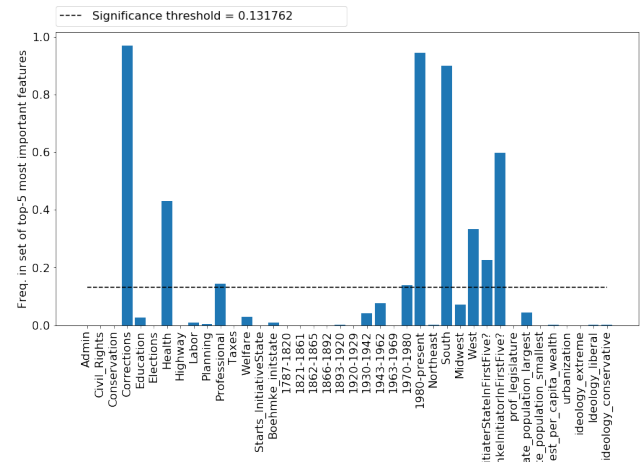}
\caption{\textbf{The frequency with which each feature was in the set of top-5 highest important scores for each of 1000 trials, for predicting whether a state-level policy ultimately became a national action.}
High frequency covariates were often near the root of the decision trees in a random forest model. The significance threshold is set by summing the probability of a given variable being chosen by chance to be in the set of top-5 variables: $(1/40) + (1/39) + (1/38) + (1/37) + (1/36) = 0.132$. Significant variables must be greater than the significance threshold.}
\label{fig:fig13S1}
\vspace{-20pt}
\end{center}
\end{figure}
% -------------------------------------------------------

\newpage

\bibliographystyle{ScienceAdvances}
\bibliography{bmc_article}
\end{document}